\pdfoutput=1
\documentclass{article}
\usepackage{spconf}
\usepackage{amsmath,graphicx}

\usepackage{newtxtext,newtxmath}
\usepackage{tabularx}
\usepackage{booktabs}
\usepackage{amsfonts}
\usepackage{cite}
\usepackage{bm}
\usepackage{url}

\usepackage{hyperref}
\hypersetup{
    colorlinks = false,
}

\title{Residual Adapters for Few-Shot Text-to-Speech Speaker Adaptation}
\name{Nobuyuki~Morioka$^{\,1}$,~~Heiga~Zen$^{\,1}$,~~Nanxin~Chen$^{\,2}$,~~Yu~Zhang$^{\,2}$,~~Yifan~Ding$^{\,1}$}
\address{$^{1\,}$Google Research, Japan \quad $^{2\,}$Google Research, USA \\[1mm]\texttt{\small \{nmorioka,heigazen,nanxinchen,ngyuzh,dyf\}@google.com}}

\copyrightnotice{\copyright\ 2022 IEEE. }
 
\begin{document}
\ninept
\maketitle
\begin{abstract}
Adapting a neural text-to-speech (TTS) model to a target speaker typically involves fine-tuning most if not all of the parameters of a pretrained multi-speaker backbone model.
However, serving hundreds of fine-tuned neural TTS models is expensive as each of them requires significant footprint and separate computational resources (e.g., accelerators, memory).
To scale speaker adapted neural TTS voices to hundreds of speakers while preserving the naturalness and speaker similarity, this paper proposes a parameter-efficient few-shot speaker adaptation, where the backbone model is augmented with trainable lightweight modules called residual adapters.
This architecture allows the backbone model to be shared across different target speakers.
Experimental results show that the proposed approach can achieve competitive naturalness and speaker similarity compared to the full fine-tuning approaches, while requiring only $\sim$0.1\% of the backbone model parameters for each speaker.
\end{abstract}
\section{Introduction}
\label{sec:intro}
Few-shot speaker adaptation continues to be an important area of research in
TTS because of growing interests in many different
commercial custom voice applications, e.g., personalized voice assistants,
in low resource scenarios \cite{chen2021adaspeech}.
Typical few-shot speaker adaptation approaches involve adapting a pretrained
multi-speaker backbone acoustic model to a target speaker by fine-tuning
all or a partial set of model parameters.
However, scaling such approaches to thousands of speakers is difficult, as the model serving cost normally grows linearly in the number of supported target speakers \cite{biadsy2022scalable}.
Furthermore, in some cases, supplementary data is needed during the fine-tuning phase, e.g., training data used for pretraining the backbone model to avoid over-fitting and/or catastrophic forgetting \cite{boffintts,Hemati2021continual,neekhara2021adapting}.

This paper introduces a parameter-efficient few-shot speaker adaptation 
approach based on residual adapters \cite{rebuffi2017learning} for TTS.
Our approach is parameter-efficient, compared to fine-tuning, as for each new speaker it only adds new parameters for an adapter (where the number of parameters in the adapter is small compared to the total number of parameters in the model), whereas in a fine-tuning setting, a complete model is created.
During adaptation, a pretrained multi-speaker TTS backbone model is frozen but
augmented with simple neural modules called residual adapters whose
parameters are optimized to synthesize a target speaker voice,
thereby allowing the same backbone model to be shared across many
different speakers for inference \cite{biadsy2022scalable}.
This is more flexible than the aforementioned fine-tuning approaches as the size of the layers in the backbone model is fixed, but the size of the residual adapters can be adjusted accordingly to the task at hand.
Also, as the residual adapters are usually tiny, requiring $\sim$0.1\% of additional parameters to the backbone model, they can prevent the serving cost to grow linearly in the number of target speakers.
For example, serving 1000 target speakers requires 1 backbone model and 1000 residual adapters which increases the cost by 2$\times$ instead of 1000$\times$.
Since the backbone model is frozen during the speaker adaptation phase, the quality of the speakers included in the backbone model is unaffected by design.

The main contributions of this paper are as follows.
\begin{itemize}
\item We propose a novel parameter-efficient approach to few-shot TTS speaker adaptation with residual adapters, and demonstrate its efficacy by showing that a few minutes of target speaker data is enough to achieve high speaker similarity.
\item We provide ablation studies that investigate the effect of residual adapters with different settings (e.g., size, location).
\end{itemize}
Audio samples are available online \cite{samples}.


\section{Related Work}
\label{sec:related}
Many previous studies on few-shot speaker adaptation are based on 
adapting pretrained multi-speaker TTS models in one way or another.
Although fine-tuning the entire model \cite{chen2018sample,arik2018neural} or
parts of the model \cite{boffintts,zhang2020adadurian} to
a new speaker has been shown to produce high-quality models in
terms of both naturalness and speaker similarity, it requires
many parameters to be updated, hence faces a scaling challenge.
AdaSpeech \cite{chen2021adaspeech} achieves parameter-efficiency
by modulating the conditional layer normalization in the decoder
with a few additional parameters for each new speaker.
In a similar spirit, our proposed approach leverages residual adapters to achieve parameter-efficiency and is more general as residual adapters can be inserted anywhere in a pretrained backbone model.
Voice Filter \cite{gabrys2022voice} decoupled the few-shot speaker
adaptation problem into speech content and speaker identity generation
tasks by introducing a trainable voice conversion module as a
post-processing step on the mel-spectrogram output of the backbone model.
To some extent, the use of residual adapters in the proposed approach can be
considered as speaker identity generation as the backbone model is 
completely frozen during adaptation.
While several works on zero-shot speaker adaptation have shown
promising results \cite{arik2018neural,jia2018transfer,cooper2020zero,yan2021adaspeech,yourtts,adaspeech4},
the focus of the proposed approach is on few-shot learning.

Speaker adaptation is more broadly categorized as task adaptation or 
transfer learning. This general problem has also been studied extensively
in other domains from which the proposed approach draws inspirations.
In particular, residual adapters have demonstrated remarkable successes in
natural language processing \cite{houlsby2019parameter},
machine translation \cite{bapna2019simple,pham2020study},
dialog systems \cite{madotto2020continual},
speech translation \cite{le2021lightweight}, 
automatic speech recognition \cite{tomanek2021residual}, and
computer vision \cite{rebuffi2017learning}.
Recently, Submodels \cite{biadsy2022scalable}, based on residual adapters, were introduced as a modular framework to address scaling challenges in model training and serving to support many customers in speech model personalization under a fixed computational resource.
Such challenges are highly relevant for commercial custom voice applications as well.
To the best of our knowledge, residual adapters have not yet been applied to scale speaker adaptation for TTS.

\section{Few-Shot Speaker Adaptation by Residual Adapters}
\label{sec:method}
At a high-level, we first pretrain a multi-speaker backbone TTS model on a high-quality parallel corpus and then adapt the backbone model to a new speaker voice by optimizing residual adapters that augment the backbone model.

\subsection{Backbone model training}
\begin{figure}
  \centering
  \includegraphics[width=0.85\linewidth]{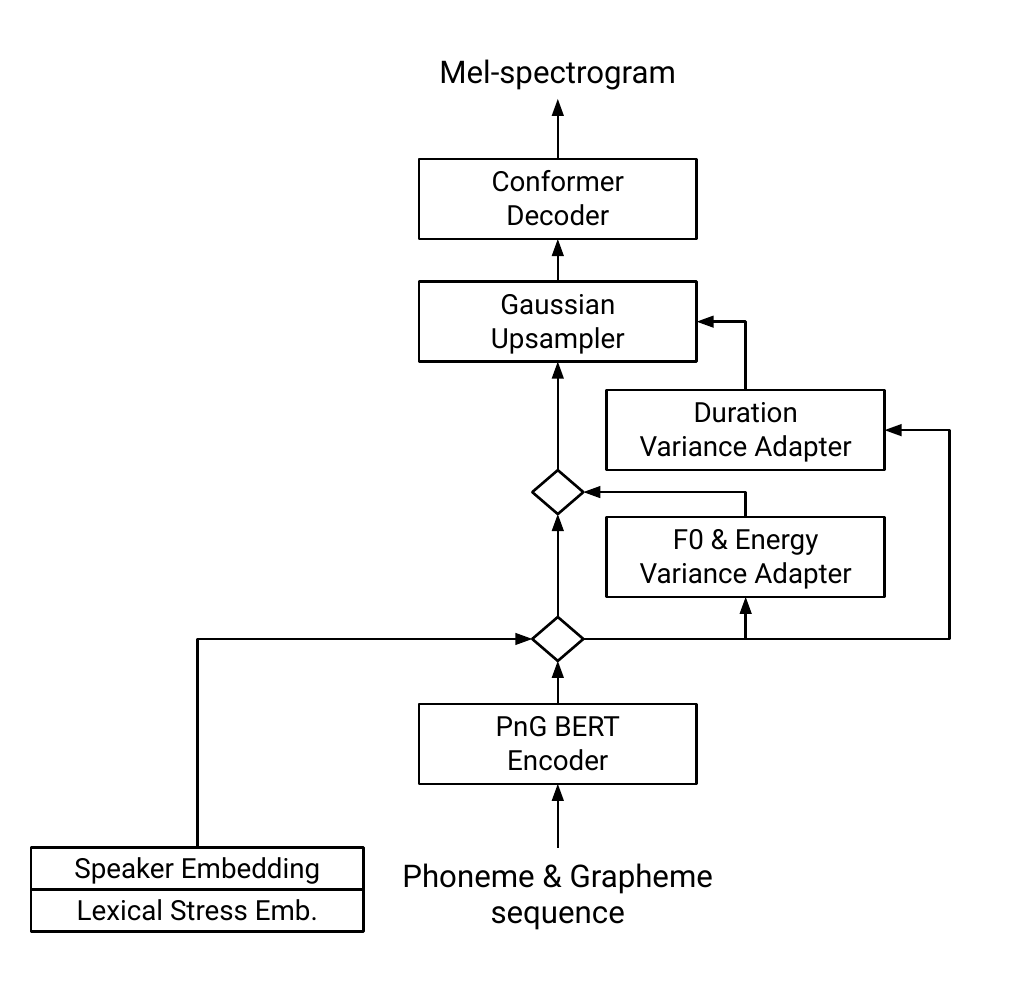}
  \caption{An illustration of a PnG NAT TTS model with phoneme-level variance adapters and a non-autoregressive Conformer-based decoder. Residual adapters can be inserted into the variance adapters and the decoder.}
  \label{fig:pngnat}
\end{figure}
The architecture of the backbone model is shown in Fig.~\ref{fig:pngnat}.
It is a non-autoregressive variant of PnG NAT (Phoneme-and-Grapheme Non-Attentive Tacotron)
\cite{jia2021png,shen2020non} that replaces an autoregressive LSTM-based decoder \cite{shen2018natural} with a non-autoregressive Conformer-based decoder \cite{gulati2020conformer,liu2021delightfultts}.
Instead of LSTM, Conformer is more straightforward to insert and experiment residual adapters, hence the change.

The model takes in both phoneme and grapheme sequences as inputs and
outputs 128-bin mel-spectrograms with a 50ms frame window and a 12.5ms frame step.
It consists of 6 stacked PnG BERT (Bidirectional Encoder Representations from Transformers) encoders, 6 stacked Conformer decoders,
a duration-based Gaussian upsampler \cite{shen2020non} and convolution-based variance adapters \cite{ren2020fastspeech2} predicting $\log$ duration, $\log F_0$ and energy at the phoneme level like FastPitch \cite{FastPitch}.
The encoder output is concatenated with an utterance-level speaker embedding that are broadcasted to the phoneme level and per-phoneme lexical stress embeddings before the variance adapters as conditioning.
Prior to the upsampler, $\log F_0$ and energy are concatenated to the encoder output with the speaker and lexical stress embeddings.

As described in \cite{jia2021png}, the PnG BERT encoder model is first pretrained on a large text corpus such as Wikipedia with the MLM (masked language model) objective and then used to initialize the encoder of the PnG NAT backbone model.
To prevent overfitting, both phoneme and grapheme embeddings and the lower 4 encoder layers are frozen while training the backbone model.

The backbone model uses teacher-forced training. Specifically, log-duration,
$\log F_0$ and energy are teacher-forced while computing a $L_1 + L_2$ loss between
the predicted and the target mel-spectrograms.
The variance adapters use $L_2$ loss for log-duration and $L_1 + L_2$ loss for
$\log F_0$ and energy.
To extract target durations, we use an external, flatstart trained, speaker-dependent
HMM-based aligner \cite{talkin1994aligner}.
For $\log F_0$ extraction, we use the RAPT algorithm \cite{talkin1995robust}.

Finally, to convert the mel-spectrogram output into an audio waveform, a speaker-independent universal pretrained WaveRNN-based neural vocoder \cite{kalchbrenner2018efficient} is used.

\subsection{Speaker adaptation by residual adapters}
\begin{figure}
  \centering
  \includegraphics[width=0.85\linewidth]{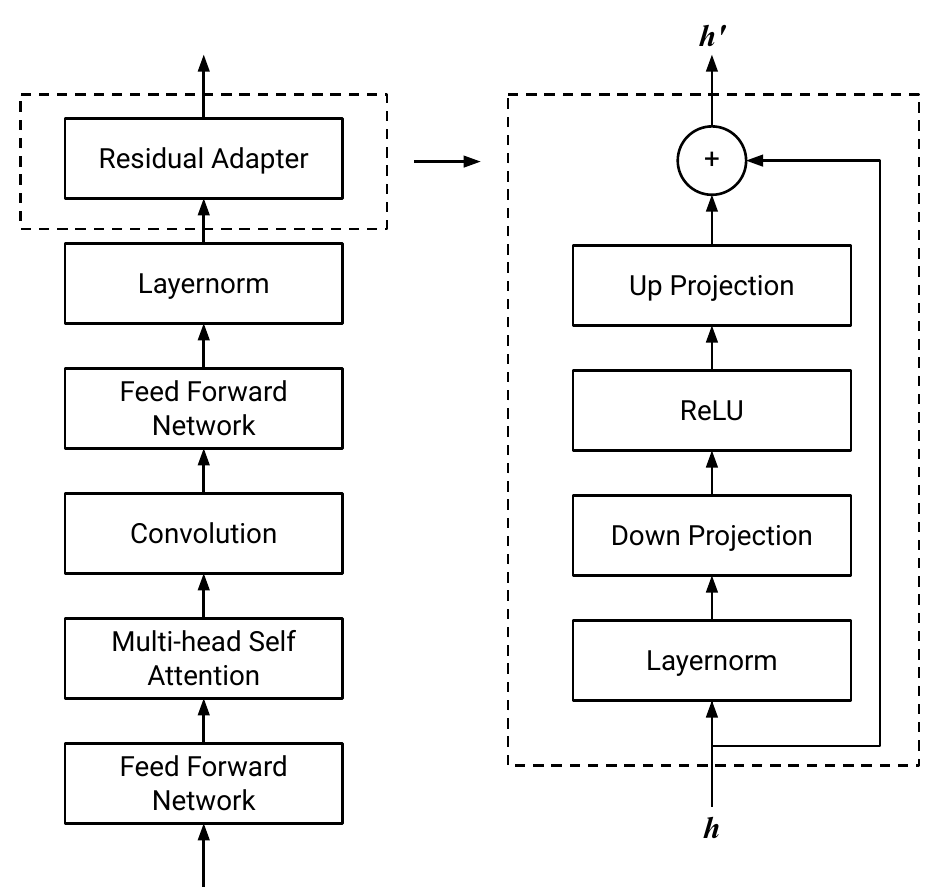}
  \caption{Left: An illustration of a residual adapter being inserted into a Conformer-based decoder layer. Right: The architecture of the residual adapter used.}
  \label{fig:resadpt}
\end{figure}
To adapt the pretrained backbone model to a new target speaker,
we employ residual adapters \cite{bapna2019simple}, 
lightweight neural modules which are inserted between layers of the backbone model.
As illustrated by Fig.~\ref{fig:resadpt},
each adapter consists of applying layer normalization \cite{layernorm} to a $d$ dimensional input vector $\bm{h} \in \mathbb{R}^{d}$
followed by down projecting to a bottleneck dimension $r$ with $\bm{W}_{\text{down}} \in \mathbb{R}^{d\times r}$, a ReLU activation \cite{ReLU}, an up projection with
$\bm{W}_{\text{up}} \in  \mathbb{R}^{r\times d}$
and a residual connection, and has the following form: 
\begin{equation}
    \bm{h}' = \bm{h} + \text{ReLU}\left(\text{LayerNorm}\left(\bm{h}\right)\bm{W}_{\text{down}}\right)\bm{W}_{\text{up}}.
\end{equation}
While the residual adapters can be inserted anywhere in the backbone model,
preliminary experiments demonstrated that they were most effective to be inserted to the Conformer layers in the decoder.
In addition to optimizing the residual adapters,
the speaker embedding for the target speaker is optimized such that the variance adapters are
conditioned properly by the learned target speaker embedding.
The entire backbone model is frozen including moving mean and variance updates in
batch normalization \cite{batchnorm}.

\section{Experiments}
\label{sec:exp}

\subsection{Experimental conditions}
To evaluate the efficacy of residual adapters for TTS speaker adaptation tasks, we conducted experiments using 1 male and 1 female US English target speakers with a neutral speaking style.
These speakers were held out from model training.
In all experiments, we adapted from the same multi-speaker non-autoregressive PnG NAT backbone model.
The PnG BERT encoders were first pretrained following the same setup described in \cite{jia2021png}.
We used subwords tokenized by SentencePiece \cite{kudo2018sentencepiece} as graphemes and obtained phonemes using a proprietary text normalization engine \cite{kestrel}.
Then, the backbone model was initialized from the pretrained PnG BERT encoders and trained on a proprietary dataset containing 243~hours of US English speech from 31 professional speakers, downsampled to 24~kHz.
By gender, the dataset had 151~hours and 92~hours of speech from 16 female and 15 male speakers respectively.
The Adam optimizer \cite{kingma2015adam} with a Transformer learning rate schedule was used to train for 450k~steps.
The backbone model had 89M parameters.

As for the residual adapters, unless otherwise stated, they were inserted between
the 6 Conformer layers in the decoder with the bottleneck dimension $r$ set to 16.
This resulted in 0.1M trainable parameters, $\sim$0.1\% of the backbone model parameters.
We also applied dropout with a rate of 0.1 just before the residual connection.
During adaptation, both the residual adapters and the speaker embedding were trained on 30 minutes of 
the target speaker data with a batch size of 256 for 50k~steps using the Adam optimizer
with a cosine learning schedule \cite{SGDR} with initial and final learning rates of $10^{-3}$
and $10^{-5}$ and decay steps of 40k.

The performance was evaluated by conducting 5-scale subjective mean opinion score (MOS) tests and side-by-side (SxS) preference tests.
Each MOS and SxS test was conducted using $\sim$1000 sample utterances.
One subject could evaluate a maximum of six pairs in the preference test and six stimuli in the MOS test.
In the MOS tests, after listening to a stimulus, the subjects were asked to rate the
naturalness of the stimulus in a 5-scale Likert scale (1: Bad, 2: Poor, 3:
Fair, 4: Good, 5: Excellent).
In the SxS tests, after listening to each pair of samples, the subjects were asked to rate the preference in naturalness in a 7-scale (-3 to 3) score.
Negative SxS scores indicate that a reference model sounded more natural than the candidate one.  
We used the above-mentioned setup of the residual adapters as the reference model in all the preference tests.
We also ran 2-scale (0: Different Speakers, 1: Same Speaker) speaker similarity tests comparing how synthetic samples were similar to original recordings.
Since the amount of the original recordings available for the speaker similarity tests was limited, the number of ratings per sample was set to 5.
Each speaker similarity test had a total of $\sim$200 ratings.
Finally, we computed objective speaker similarity metric using mean cosine similarity of d-vectors \cite{dvector,variani2014deep} between original recordings and synthesized samples.
Larger scores are better in these metrics.

\begin{table*}[t!]
\centering
\caption{Effect of training residual adapters with and without backbone training data. Both setups use 30 minutes of the target speaker data.}
\vspace{0.5mm}
\label{tbl:basetraindata}
\begin{tabular}{lcccccccc}
\toprule
& \multicolumn{4}{c}{Female Speaker} & \multicolumn{4}{c}{Male Speaker} \\
\cmidrule(lr){2-5} \cmidrule(lr){6-9}
Data & MOS & SxS & Similarity & Cosine Sim. & MOS & SxS & Similarity & Cosine Sim. \\
\midrule
30 min & 4.46$\pm$0.05 & Reference & 0.86$\pm$0.10 & 0.78$\pm$0.02
                  & 4.51$\pm$0.05 & Reference & 0.93$\pm$0.07 & 0.75$\pm$0.02 \\
30 min + backbone & 4.49$\pm$0.04 & 0.00$\pm$0.06 & 0.89$\pm$0.09 & 0.77$\pm$0.02
                   & 4.45$\pm$0.05 & 0.00$\pm$0.05 & 0.77$\pm$0.12 & 0.73$\pm$0.02 \\
\bottomrule
\end{tabular}
\vspace{1mm}
\centering
\caption{Effect of inserting different size residual adapters into the Conformer decoder and the variance adapters.
$r$ denotes the bottleneck dimensions of the residual adapters. The percent of additional parameters relative to the entire model for each setup is also shown.}
\vspace{0.5mm}
\label{tbl:sizeloc}
\setlength{\tabcolsep}{0.43em}
\begin{tabular}{llccccccccc}
\toprule
& & & \multicolumn{4}{c}{Female Speaker} & \multicolumn{4}{c}{Male Speaker} \\
\cmidrule(lr){4-7} \cmidrule(lr){8-11}
Decoder & Var. Adpt. & Params & MOS & SxS & Similarity & Cosine Sim.
                                 & MOS & SxS & Similarity & Cosine Sim. \\
\midrule
$r=16$ & -- & 0.12\% & 4.46$\pm$0.05 & Reference & 0.86$\pm$0.10 & 0.78$\pm$0.02
                     & 4.51$\pm$0.05 & Reference & 0.93$\pm$0.07 & 0.75$\pm$0.02 \\
$r=128$ & -- & 0.89\% & 4.48$\pm$0.05 & 0.00$\pm$0.05 & 0.84$\pm$0.11 & 0.80$\pm$0.02
                      & 4.50$\pm$0.04 & \textbf{-0.05$\pm$0.05} & 0.97$\pm$0.04 & 0.76$\pm$0.02 \\
$r=16$ & $r=8$ & 0.16\% & 4.39$\pm$0.05 & \textbf{-0.17$\pm$0.08} & 0.89$\pm$0.09 & 0.80$\pm$0.02 
                        & 4.28$\pm$0.05 & \textbf{-1.04$\pm$0.10} & 1.00$\pm$0.00 & 0.77$\pm$0.02 \\
$r=128$ & $r=64$ & 1.20\% & 4.18$\pm$0.06 & \textbf{-1.02$\pm$0.10} & 0.86$\pm$0.10 & 0.80$\pm$0.02
                          & 4.34$\pm$0.05 & \textbf{-0.46$\pm$0.08} & 0.95$\pm$0.06 & 0.79$\pm$0.02\\
\bottomrule
\end{tabular}
\vspace{1mm}
\centering
\caption{Effect of varying the amount of target speaker data on residual adapters.}
\vspace{0.5mm}
\label{tbl:dataamount}
\begin{tabular}{lcccccccc}
\toprule
& \multicolumn{4}{c}{Female Speaker} & \multicolumn{4}{c}{Male Speaker} \\
\cmidrule(lr){2-5} \cmidrule(lr){6-9}
Data & MOS & SxS & Similarity & Cosine Sim. & MOS & SxS & Similarity & Cosine Sim. \\
\midrule
30 min & 4.46$\pm$0.05 & Reference & 0.86$\pm$0.10 & 0.78$\pm$0.02
       & 4.51$\pm$0.05 & Reference & 0.93$\pm$0.07 & 0.75$\pm$0.02 \\
5 min & 4.45$\pm$0.05 & -0.01$\pm$0.05 & 0.57$\pm$0.15 & 0.77$\pm$0.02
      & 4.49$\pm$0.04 & -0.04$\pm$0.05 & 0.93$\pm$0.07 & 0.75$\pm$0.02 \\
1 min & 4.42$\pm$0.05 & \textbf{-0.08$\pm$0.07} & 0.90$\pm$0.12 & 0.77$\pm$0.03
      & 4.46$\pm$0.05 & \textbf{-0.12$\pm$0.07} & 0.95$\pm$0.08 & 0.73$\pm$0.03 \\
\bottomrule
\end{tabular}
\vspace{1mm}
\centering
\caption{Comparison against full fine-tuning baselines trained on different training data setups.}
\vspace{0.5mm}
\label{tbl:baseline}
\begin{tabular}{lcccccccc}
\toprule
& \multicolumn{4}{c}{Female Speaker} & \multicolumn{4}{c}{Male Speaker} \\
\cmidrule(lr){2-5} \cmidrule(lr){6-9}
& MOS & SxS & Similarity & Cosine Sim. & MOS & SxS & Similarity & Cosine Sim. \\ 
\midrule
Residual Adapters & 4.46$\pm$0.05 & Reference & 0.86$\pm$0.10 & 0.78$\pm$0.02
                  & 4.51$\pm$0.05 & Reference & 0.93$\pm$0.07 & 0.75$\pm$0.02 \\
\midrule
\multicolumn{5}{l}{Fine-tuning w/ backbone data} \\
\multicolumn{1}{l}{\kern0.5em30 min}
    & 4.49$\pm$0.05 & -0.01$\pm$0.07 & 0.84$\pm$0.11 & 0.80$\pm$0.02
    & 4.49$\pm$0.05 & -0.04$\pm$0.07 & 0.95$\pm$0.06 & 0.77$\pm$0.02 \\
\multicolumn{5}{l}{Fine-tuning w/o backbone data} \\
\multicolumn{1}{l}{\kern0.5em30 min}
    & 4.35$\pm$0.05 & \textbf{-0.33$\pm$0.09} & 0.86$\pm$0.10 & 0.80$\pm$0.02
    & 4.30$\pm$0.05 & \textbf{-0.47$\pm$0.09} & 0.93$\pm$0.07 & 0.79$\pm$0.02 \\
\multicolumn{1}{l}{\kern0.5em5 min}
    & 4.28$\pm$0.05 & \textbf{-0.54$\pm$0.09} & 0.78$\pm$0.12 & 0.78$\pm$0.02
    & 4.30$\pm$0.05 & \textbf{-0.57$\pm$0.09} & 0.93$\pm$0.07 & 0.78$\pm$0.02 \\
\multicolumn{1}{l}{\kern0.5em1 min}
    & 4.13$\pm$0.06 & \textbf{-0.77$\pm$0.09} & 0.57$\pm$0.21 & 0.75$\pm$0.03
    & 4.06$\pm$0.06 & \textbf{-0.91$\pm$0.10} & 0.56$\pm$0.20 & 0.75$\pm$0.03 \\
\bottomrule
\end{tabular}
\end{table*}
\begin{table}[t!]
\centering
\caption{Cosine similarity comparison against zero-shot d-vector and few-shot speaker embedding only fine-tuning baselines.}
\vspace{0.5mm}
\label{tbl:zeroshot}
\setlength{\tabcolsep}{0.45em}
\begin{tabular}{llcc}
\toprule
& Data & Female Speaker & Male Speaker \\
\midrule
Residual Adapters & 1 min & 0.77$\pm$0.03 & 0.73$\pm$0.03 \\
                  & 1 sample & 0.74$\pm$0.03 & 0.68$\pm$0.03 \\
\midrule
Zero-shot d-vector & 1 min & 0.67$\pm$0.03 & 0.57$\pm$0.03 \\
                   & 1 sample & 0.61$\pm$0.03 & 0.58$\pm$0.02 \\
\midrule
Fine-tuning speaker & 1 min & 0.64$\pm$0.04 & 0.48$\pm$0.02 \\
\, embedding only   & 1 sample & 0.30$\pm$0.03 & 0.50$\pm$0.02 \\
\bottomrule
\end{tabular}
\end{table}

\subsection{Ablation studies}

We first present ablation studies to understand the characteristics of
residual adapters for TTS speaker adaptation.

Table~\ref{tbl:basetraindata} shows the performance of including and excluding the backbone training data at the speaker adaptation stage.
The mix ratio of the backbone training data and 30 minutes of the target speaker data used was 99:1.
There was no significant naturalness difference between the two setups.
As the residual adapters had less parameters to update, it was able to mitigate overfitting and/or catastrophic forgetting without the backbone data.
For the speaker similarity tests, while the confidence intervals overlapped, the male speaker had a slight drop when training with the backbone data.
The drop may be due to more female speech in the backbone data.

Table~\ref{tbl:sizeloc} compares residual adapters with different sizes inserted into the Conformer decoder and the variance adapters.
We noticed a slight regression in naturalness with larger residual adapters (1st row vs. 2nd row) likely due to overfitting, though no significant difference in speaker similarity.
Surprisingly, adding the residual adapters in the variance adapters resulted in significantly worse naturalness (1st row vs. 3rd row).
In one of the speakers, the naturalness gap further widened for bigger residual adapters (1st row vs. 4th row).
One possible explanation is that updating the speaker embedding was enough to predict the variance information of the target speakers.
The rater comments for the speech produced by the model with residual adapters applied to the variance adapters included slow speed, incorrect pitch, aggressive tone.
It would be interesting to investigate if target-speaker-specific characteristics can be captured solely by the residual adapters, using a backbone model without a speaker embedding in future work.

Table~\ref{tbl:dataamount} shows the effect of varying the amount of the target speaker data.
There was no significant difference in naturalness between the models trained on 30 minutes and 5 minutes of the target speaker data (1st row vs. 2nd row).
A significant drop in speaker similarity for the female speaker might be due to a rating fluctuation, as the cosine similarities show that the synthesized samples of the female speaker across different amounts of data were equally similar to the original recordings.
For reference, the cosine similarity between the female speaker's original recordings and the male speaker's synthesized samples was 0.22 and that between the male speaker's original recordings and the female speaker's synthesized samples was 0.20.
Also, the cosine similarities between the synthesized samples of these speakers and the original recordings of other speakers of the same gender were 0.39 for female and 0.31 for male.
While 1 minute of the target speaker data resulted in a marginal drop in quality (1st row vs. 3rd row), it demonstrated the robustness of residual adapters even when data is limited.
The rater comments on comparing samples synthesized by the models trained on 30 minutes and 1 minute included the former being slightly more natural in terms of speed and tone than the latter.

\subsection{Comparison against fine-tuning and zero-shot baselines}

Next, we evaluated the proposed approach with full fine-tuning baselines trained on various training data setups.
The baselines were fine-tuned from the same backbone model and all model parameters except for the encoder were updated for 50k steps.
As for the learning rate, the speaker embedding used the same cosine learning rate schedule and the rest used a constant learning rate of $10^{-5}$.
Table~\ref{tbl:baseline} shows that while the fine-tuning baseline using both the backbone training data and 30 minutes of the target speaker data performed competitively against the proposed approach, naturalness quickly degraded as we excluded the backbone training data and decreased  the amount of the target speaker data.
Moreover, with 1 minute of the target speaker data speaker, the baseline could not achieve high speaker similarity.
In contrast, as shown in Table~\ref{tbl:dataamount}, the proposed approach was able to maintain its quality even with 1 minute of the target speaker data and being parameter efficient at the same time. 

As a zero-shot TTS baseline, we trained a backbone model that used d-vector \cite{dvector,variani2014deep} instead of the speaker embedding as conditioning in a similar way to \cite{jia2018transfer}.
During inference, we used either a d-vector computed from 1 utterance ($\sim$7 seconds) or a mean d-vector averaged over randomly sampled 10 utterances ($\sim$1 minute) as speaker conditioning.
We also trained another few-shot fine-tuning baseline that updated the speaker embedding only.
Table~\ref{tbl:zeroshot} shows that the proposed approach can still achieve better objective speaker similarity than the baselines with such small data.

\section{Conclusion}
\label{sec:conclusion}
This paper introduced a parameter-efficient approach to few-shot TTS
speaker adaptation via residual adapters.
The proposed approach achieved competitive or better performance in both naturalness
and speaker similarity compared to the fine-tuning and zero-shot speaker adaptation baselines.
Since the backbone model is frozen during adaptation, it can be shared across different speakers.
This architecture can reduce serving cost for custom voice applications and allows it to scale hundreds of speakers.
This can be further reduced to a constant memory footprint with Submodels \cite{biadsy2022scalable} that load and swap out different residual adapters on the fly.

Future work includes an evaluation of the proposed approach for target speakers with different speaking styles and in different languages.
This paper has shown one successful application of residual adapters in TTS.
It would be interesting to apply residual adapters to other use cases such as expanding an existing multilingual TTS model with additional languages to avoid training from scratch while maintaining the quality of existing languages.


\vfill\pagebreak
\newpage
\bibliographystyle{IEEEbib}
\footnotesize 
\bibliography{main}

\end{document}